# Physics Webpages Create Barriers to Participation for People with Disabilities: Five Steps to Increase Digital Accessibility

Erin Scanlon[1], Zachary W. Taylor[2], and Jacquelyn J. Chini[1*]

[1] Department of Physics, University of Central Florida, Orlando FL

[2] Department of Educational Leadership and Policy, University of Texas at Austin, Austin TX

* Corresponding author

**ABSTRACT**

**Background:** While there have been numerous calls to increase the participation of people with disabilities in STEM, many postsecondary institutions are not equipped to support students with disabilities. We examined the digital accessibility of 139 webpages from 73 postsecondary institutions in the United States that contained information about the undergraduate physics curriculum and graduate research programs. We selected these webpages as they are common entry points for students interested in pursuing a physics degree. We used Tenon[TM] and Mac OS X's Voiceover software to assess the accessibility of these webpages as measured by alignment with the Web Content Accessibility Guidelines (WCAG) 2.0.

**Results:** We found only one webpage was accessible for students with disabilities. Five specific error-types accounted for 95.5% of all errors. The five most common errors were related to information, structure and relationships of content (1.3.1 Level A; 38.2%); text alternatives for non-text content (1.1.1 Level A; 24.6%); information about link purpose (2.4.4 Level A; 14.7%); capability to resize text (1.4.4 Level AA; 10.1%); and information about the name, role and value of user interface components (4.1.2 Level A; 7.9%).

**Conclusions:** We present and describe the five common accessibility errors we identified in the webpages in our sample, suggest solutions for these errors, and provide implications for students with disabilities, instructors and staff, institutional administration, and the broader physics community.

**I. INTRODUCTION**

There have been many calls to increase the participation of people with disabilities in the science, technology, engineering, and mathematics (STEM) workforce. For example, in the United States the National Science Foundation (NSF; 2011) states, "Tomorrow's STEM workforce must draw on the talents and interests of all sectors of the nation's diverse population. NSF will intensify our efforts to expand participation in the STEM workforce by currently underrepresented segments of the population—women, minorities, and persons with disabilities." (p. 3) However, to enter the STEM workforce, students need to matriculate through STEM degree programs and students with disabilities are underrepresented in postsecondary STEM programs (NCSES, 2015).

People with disabilities demonstrate high interest in STEM during the transition from high school to college, but their representation in STEM decreases throughout postsecondary education and into the workforce. Looking at statistics in the United States, students with disabilities made up 11% of the undergraduate student population in 2012 (NCSES, 2015) and represent 19.8% in 2017 (NSF, 2019) when measured using an expanded definition of executive function disorders (i.e., questions related to difficulties concentrating, remembering, or deciding;

https://nces.ed.gov/surveys/npsas/pdf/N16_Data_Info.pdf). Also, undergraduate students with and without disabilities enroll in science and engineering majors at similar rates (28.0% and 27.6%, respectively; NSF, 2019). However, in 2017 only 335 doctoral degrees in physical sciences were awarded to people with disabilities (7.4% of all physical sciences degrees; NSF, 2019), and people with disabilities make up 9.3% of employed physical scientists with disabilities. It makes sense that the number of employed physical scientists has a higher representation of people with disabilities because people can gain disabilities as they age. For example, 8.4% of employed people with disabilities in STEM were diagnosed at birth, 11.4% were diagnosed at age 20-29 (i.e., typical college age), and 21.9% were diagnosed at age 50-75 (NSF, 2019).

There are a myriad of factors that may contribute to people with disabilities ceasing participation in STEM, including: different academic preparation than their peers (whether caused by restrictive Individual Education Plans or their own interests; Sparks & Lovett, 2009), lack of curricular materials designed to support students with disabilities (Scanlon, Legron-Rodriguez, Schreffler, Ibadlit, Vasquez, & Chini, 2018; Scanlon, Schreffler, James, Vasquez, & Chini, 2018), lower sense of belonging in the postsecondary community (Kurth & Mellard, 2006), lack of faculty preparation to support students with disabilities (Scott, 2009; Thompson, Bethea, & Turner, 1997; Moriarty, 2007; Love et al., 2014), and motivational factors such as lower expectations of success in science and mathematics (Bittinger, 2018).

Another contributing factor to the decrease in participation by people with disabilities is related to the accessibility of the information related to academic programs. Kane, Shulman, Shockley, and Ladner (2007) state: "University [web]sites that are not accessible may exclude people with disabilities from participation in educational, social and professional activities" (p.

148). In 2011, 93% of institutions of higher education had a main website where information about the institution was housed and the National Center of Education Statistics (2011) reported that only 24% of these institutions follow established accessibility guidelines (Raue & Lewis, 2011). This means that at least some of the information presented on these institutions' websites is not accessible to some people; this inaccessible information can be a significant barrier to participation for people with disabilities. If prospective students do not have access to the information about an institution, they may be less likely to enroll at that institution (Burdett, 2013; Daun-Barnett & Das, 2013). The inaccessibility of postsecondary institutions' websites creates a barrier to participation for people with disabilities.

The same is true for postsecondary physics websites; if a program's webpages are inaccessible, it makes sense that people with disabilities would be less likely to enroll in that program. The purpose of this study was to conduct a case study examination of common digital entry points to undergraduate and graduate physics programs in the United States and to assess their accessibility. We chose to examine undergraduate curriculum webpages and graduate research webpages as they are common places students look for information about a physics program. Below, we summarize the basic principles of web accessibility and the related requirements (under U.S. law and policy) and responsibilities of organizations of higher education.

**A. Web accessibility principles**

The World Wide Web Consortium (W3C) is an international community that created a set of universally accepted requirements in the Web Content Accessibility Guidelines (WCAG). The basic principles of web accessibility fall into four categories to ensure web content is:

perceivable, operable, understandable, and robust (Accessibility Principles, 2019). *Perceivable* web content provides multiple ways for users to access (e.g., visual, auditory) and customize (e.g., enlarge, change colors) information. *Operable* web content provides multiple ways to control (e.g., mouse, keyboard) and navigate (e.g., search boxes, site maps) webpages. *Understandable* web content is readable (e.g., identifies language), understandable (e.g., uses clear language), and helps users to avoid mistakes (e.g., provides error messages). *Robust* web content is compatible with multiple browsers and assistive technologies (e.g., alternate keyboards, screen readers). Further descriptions and examples of accessibility principles and practices for web content are provided in the appendix; this is not an exhaustive list.

Each principle is composed of guidelines which are each delineated by more specific "success criteria", which are rated on a three-level scale (A, AA, and AAA). The individual success criteria were assigned a level based on a number of interacting issues, including the impact of the success criterion on the accessibility of the content (e.g., a success criterion is essential if the content will be inaccessible even with assistive technology if the criterion is not met) and the difficulty of implementing the success criterion (e.g., the criterion requires skills that are reasonably achievable by content creators; Understanding Conformance, 2016). For example, the entries in Table A1 for guidelines 1.2 (Captions), 1.4 (Distinguishable), and 3.3 (Input Assistance) demonstrate success criteria at all three levels of conformance.

**B. Accessibility requirements and responsibilities for higher education**

Inaccessible webpages limit access for individuals with disabilities. For example, poorly designed webpages create barriers for individuals with visual impairments (who may interact with the content via a screen reader), mobility impairments (who may navigate the webpage with a keyboard rather than a mouse), and attention disorders (who may use multiple means to navigate content efficiently).

**i. United States Laws**

Section 508 of the Rehabilitation Act describes the online accessibility standards that U.S. federal agencies, contractors, and employers must meet (US General Services Administration). As of January 18, 2018, Section 508 requires that all U.S. federal information and communication technology services are compliant with WCAG 2.0 Level A and Level AA criteria. While American universities differ in their interpretations about whether Section 508 applies to them [EduCase], all Title IV-participating institutions (those that award federal student loans) must maintain accessible websites under Section 504. Since Section 504 (U.S. General Services Administration) does not explicate accessibility standards, Section 508 is generally accepted as the baseline for the federal government's expectations for accessibility [Inside Higher Ed].

**ii. International Laws**

There are many international laws that also cite the WCAG 2.0 standards. For example, Australia (Disability Discrimination Act, 1992), Canada (Policy on Communications and Federal Identity, 2016), European Union (Web and Mobile Accessibility Directive, 2016), India (Guidelines for Indian Government Websites, 2009), and Israel (Equal Rights of Persons with Disabilities Act 1998) all have accessibility laws that suggest or mandate WCAG compliance (World Wide Web Consortium, 2019).

**C. Previous digital accessibility studies in higher education**

In last few decades, there have been multiple studies assessing the digital accessibility of higher education webpages in the United States (Thompson, Burgstahler, & Comden, 2003; Floyd & Santiago, 2007; Thompson, Burgstahler, & Moore, 2010; Forgione-Barkas, 2012; Kimmons, 2017) and abroad (United Kingdom: Sloan, Gregor, Booth, & Gibson, 2002; Malaysia: Aziz, Isa, & Nordin, 2010; Portugal: Gonçalves, Martins, Pereira, & Cota, 2013; Albania: Ferati, Mripa, & Bunjaku, 2016; Kyrgyzstan, Azerbaijan, Kazakhstan, and Turkey: Ismailova & Inal, 2018; Ocenia and Arab countries: Alahmadi & Drew, 2017). These investigations have spanned all levels of institutions of higher education including community colleges (Erickson, Trerise, Lee, VanLooy, Knowlton, & Bruyere, 2013), research institutions (McGough, 2016), and "top" international institutions (Kane, Shulman, Shockley, & Ladner, 2004).

These studies indicate that there are many accessibility errors present in institutional higher education webpages (i.e., there are many inaccessible components to higher education webpages). For example, Harper and DeWaters (2008) found that most university homepages in their study were not compliant with the WCAG 2.0 guidelines, and Forgione-Barkas (2012) found most errors in her study occurred within the Level A conformance standards. Similarly, when Thompson, Burgstahler, and Moore (2010) examined changes in web accessibility over a five-year period, they found "although significant positive gains regarding accessibility were revealed in some areas, such as alternate text on images and coded support for navigation, even in these areas the percentage of pages that are accessible is strikingly low" (p. 113).

Kimmons (2017) recently examined the homepages and first-level subpages of all institutions of higher education in the United States for compliance with web accessibility standards and found that 71.5% of institutions' homepages contained at least one error that would make it inaccessible, with an average of 5.89 errors per page. Kimmons (2017) states "website accessibility still seems to be a systemic struggle for institutions of higher education, as evidenced by the very high error rates present across homepages and subpages" (p. 447).

Noh, Jeong, You, Moon, & Kang (2015) previously studied digital accessibility of Korean science and technology institutions of higher education and found significant inaccessibility. 68% of websites complied with Perceivable guidelines, 64.5% complied with Operable guidelines, 59.2% complied with Understandable guidelines, and 28.0% complied with Robust guidelines.

The literature base suggests that many postsecondary institutions' webpages are not accessible for students with disabilities and may pose barriers to their full participation in the educational setting. The purpose of this study was to investigate the accessibility of U.S. webpages that prospective and current students would interact with when considering participation in a physics program.

## 2. METHODOLOGY

### A. Sample

We used the Integrated Postsecondary Education Data System (IPEDS) to locate every Title-IV participating postsecondary institution in the United States and created a random sample of 400 institutions. We included 400 institutions by considering the overall population of Title IV-participating U.S. institutions (N = 6,676). We found that 74 of these 400 institutions offered physics degrees, including associate of arts (AA), associate of science (AS), bachelor of arts (BA), bachelor of science (BS), master of science (MS), and doctor of philosophy (PhD). We did not include applied physics degrees, pre-service physics teacher programs, or programs that had a physical science focus. We identified the hyperlink to the webpage with the university's physics degree requirements for each undergraduate degree offered (e.g., https://sciences.ucf.edu/physics/undergraduate/curriculum/) and the graduate research opportunities (e.g., https://sciences.ucf.edu/physics/research/). Our final sample included 74 two-year and four-year institutions and 139 hyperlinks; the sample is summarized in Tables I and II.

Table I: Institution webpage types included in our sample

| Sector | Institutions |
|---|---|
| Public - All | 47 |
| Public - Four-year | 27 |
| Public - Two-year | 20 |
| Private - All | 27 |
| Private - Four-year | 26 |
| Private - Two-year | 1 |
| Overall Total | 74 |

Table II: Program webpage types included in our sample

| Degree Type | Programs |
|---|---|
| Associate (AA and AS) | 31 |
| Bachelor of Arts | 34 |
| Bachelor of Science | 55 |
| Graduate Research (MS and PhD) | 19 |
| Overall Total | 139 |

**B. Analysis**

We used Tenon™ accessibility software to analyze each undergraduate physics curriculum webpage and graduate research webpage for web accessibility. Tenon™ is a robust web accessibility audit software program capable of running nearly 100 total tests of web accessibility at the Level-A, Level-AA, and Level-AAA standards (Tenon LLC, 2018). As Title-IV participating institutions are not required to meet Level-AAA conformance, all Level-AAA errors discovered in this study were removed from the analysis (US General Services Administration). Tenon™ produces downloadable .csv reports which define the most prevalent web accessibility errors and the HTML location of the error. Comparative analyses of web accessibility evaluation software have found Tenon™ to be an efficient, accurate, and robust web accessibility evaluation tool (Ismail, Kuppusamy, & Nengroo, 2017; Taylor, 2018; Timbi-Sisalima et al., 2018). Tenon™ does not analyze PDF documents on webpages, so we did not analyze these webpages (pages such as http://sciences.ucf.edu/physics/wp-content/uploads/sites/116/2012/05/UNGRD-Program-of-Study.pdf).

A webpage with a single WCAG 2.0 web accessibility error is not necessarily inaccessible for all students with disabilities in its entirety (Erickson et al., 2013; Flowers et al., 2011; Hackett & Parmanto, 2005; Thompson et al., 2010). Thus, we also evaluated each webpage for web accessibility using Mac OS X's Voiceover, which is a fully-functional, robust, screen-reading assistive technology built-in to Macintosh computers and used by people with disabilities including blindness, low vision, and dyslexia. Voiceover has been found to be a reliable, efficient, and effective assistive technology and was used to add another layer of reliability beyond evaluation technologies such as Tenon™ (Edwards, 2005; Henton, 2012; Manduchi & Kurniawan, 2012).

## III. FINDINGS

### A. Overall web accessibility

Table III displays the mean, standard deviation, and high and low number of errors across the entire sample, by institution type and program type. In total, we identified 44,241 web accessibility errors across the 139 webpages. The most inaccessible webpage included 1,600 web accessibility errors; Voiceover audit of this webpage showed that every web element—all 1,600—on this webpage was inaccessible for students with disabilities. Tenon™ and Voiceover audits of all the webpages in the sample indicated that only one physics curriculum webpage was fully accessible for students with disabilities. This meant that a student with a disability could use Voiceover and interact with every web element on the webpage without any missing information. The BS in physics curriculum webpage at this small, residential public university in the Southwest had only ten Level-A errors and no Level-AA errors. All other webpages included in our sample were determined to be inaccessible for students with disabilities.

Table III: Descriptive statistics of web accessibility errors (N = 44,241) of physics curriculum webpages disaggregated by institution type and program type. Note: The cells are formatted as mean (standard deviation) with the high; low in the second line.

| Institution/Program Type | Web accessibility errors | | |
| --- | --- | --- | --- |
| | Level A | Level AA | Level A & AA* |
| All (N = 73 Institutions) | 296.3 (288.5) | 36.2 (27.2) | 322.6 (300.1) |
| | 1,556; 10 | 142; 0 | 1,600; 10 |
| Public, four-year (**N = 45) | 246.4 (181.4) | 29.3 (21.6) | 275.7 (187.1) |
| | 729; 10 | 100; 0 | 759; 10 |
| Private, four-year (N = 46) | 302.1 (315.4) | 38.1 (22.8) | 340.2 (322.1) |
| | 1,556; 29 | 120; 7 | 1,600; 48 |
| Public, two-year (N = 28) | 341.9 (353.2) | 41.2 (33.5) | 383.2 (369.3) |
| | 1,412; 38 | 124; 5 | 1,488; 43 |
| Private, two-year (N = 2) | 648.5 (525.4) | 81.5 (85.6) | 730 (610.9) |
| | 1,020; 277 | 142; 21 | 1,162; 298 |
| Associate's, AA and AS (N = 33) | 331.8 (354.9) | 41.6 (36.6) | 373.4 (377.4) |
| | 1,412; 38 | 142; 3 | 1,488; 43 |
| Bachelor's, BA and BS (N = 88) | 283.1 (260.4) | 34.3 (22.6) | 317.3 (266.2) |
| | 1,556; 10 | 120; 0 | 1,600; 10 |
| Graduate Research (N = 18) | 191.6 (166.8) | 30.3 (14.6) | 221.9 (167.5) |
| | 691; 25 | 61; 6 | 734; 57 |

*Note: Only Level A and Level AA errors were reported, as Section 508 only requires compliance with the Level A and Level AA web accessibility threshold. **Note: All subsequent

sample sizes refer to number of webpages. An individual institution may have multiple webpages in the sample.

Based on the results in Table III, we found that public, four-year institutional webpages were the most web accessible (275 errors per webpage on average), whereas public, two-year institutional webpages were the least web accessible (383 errors per page on average). There were only two private, two-year webpages included in this study, with these webpages including 730 errors per webpage on average. Across different degree types, graduate research webpages were the most web accessible (with an average of 221 errors per webpage), while associate's degree program webpages were the least web accessible (with an average of 373 errors per webpage).

**B. Most frequent web accessibility errors**

Table IV displays the most frequent errors across the four categories of WCAG 2.0. This highlights the most abundant web accessibility errors and how postsecondary students with disabilities may be unduly burdened by certain types of errors if their disability requires a specific assistive technology (e.g., an assistive technology using a keyboard to input all information).

Table IV: Descriptive statistics of web accessibility errors (N = 44,241) on physics curriculum webpages published on institutional webpages (N = 73), by error type

| Errors, by type, all institutions | # of errors | % of all errors |
|---|---|---|
| Perceivable | | |
|   Level A, 1.1.1, Non-text content | 10,887 | 24.6% |
|   Level A, 1.2.2, Captions (prerecorded) | 1 | <1% |
| Level A, 1.3.1, Information and relationships | 16,912 | 38.2% |
| Level A, 1.3.2, Meaningful sequence | 644 | 1.5% |
| Level A, 1.4.3, Contrast (minimum) | 413 | <1% |
| Level AA, 1.4.4, Resize text | 4,436 | 10.1% |
| Operable | | |
| Level A, 2.1.1, Keyboard | 616 | 1.4% |
| Level A, 2.4.1, Bypass blocks | 94 | <1% |
| Level A, 2.4.2, Page titled | 2 | <1% |
| Level A, 2.4.3, Focus order | 20 | <1% |
| Level A, 2.4.4, Link purpose (in context) | 6,514 | 14.7% |
| Level AA, 2.4.6, Headings and labels | 88 | <1% |
| Understandable | | |
| Level A, 3.1.1, Language of page | 14 | <1% |
| Level A, 3.2.1, On focus | 20 | <1% |
| Level A, 3.2.4, Consistent identification | 33 | <1% |
| Robust | | |
| Level A, 4.1.1, Parsing | 49 | <1% |
| Level A, 4.1.2, Name, role, value | 3,498 | 7.9% |
| Total | 44,241 | 100% |

Note: 1 institution only had PDF webpages which cannot be analyzed using Tenon™

The data in Table IV suggest that five Level-A and Level-AA errors were responsible for the majority of web accessibility errors in this study: non-text content (1.1.1), information and relationships (1.3.1), resize text (1.4.4), link purpose in context (2.4.4), and name, role, and value (4.1.2). Few errors were identified in the understandable category. We describe each common error below in the Discussion section.

**IV. DISCUSSION**

**A. Frequent perceivability errors and suggested corrections**

*1. Common errors*

Level-A 1.3.1 Information and relationships errors (N = 16,912) comprised the largest percentage of all errors in this study (38.2%). These errors are related to the type of text on a webpage and whether students with a wide range of disabilities, and those who use a wide range of assistive technologies, can read the content. For example, many Level-A 1.3.1 errors identified web elements with overly long passages of text (ten or more words) in bold and uppercase letters, as shown in Figure 1. Although using bold and uppercase text may be beneficial for some web users, this formatting may be difficult to read and comprehend for students with dyslexia, for example.

Figure 1: Example of Level-A 1.3.1 with the error (i.e., bold, capitalized) and without the error (i.e., not capitalized and bolded sparingly).

%%% Production note: Insert Figure 1 here %%%

Another common error in this category was embedded tables without labels. The header of a table should be structured and labeled to convey the relationship between the header and the other content in the table. Most physics curriculum webpages in this study contained embedded tables listing the required physics courses necessary for a degree; many of these embedded tables lacked header information or were labeled with uninformative text, such as "list", which does not tell the web user what is included in the table. With this type of error, people using screen readers may not know there is a table and may miss information presented in subsequent columns.

Level-A 1.1.1 Non-text content errors (N = 10,887) comprised 24.6% of all web accessibility errors in this study. Non-text content errors occur when a non-text web element (e.g., image, menu, video) is missing metadata to communicate with the assistive technology that a student with a disability may use to navigate the webpage. For instance, many webpages with this error included multiple images with the same alternative text (i.e., the text entered as part of a web element's metadata to describe the web element to the web user). If two images contain the same alternative text, the web user who is blind or has low-vision will not be able to differentiate between the two images when they hover their mouse over the image and the assistive technology reads the same alternative text aloud to the web user.

Level-A 1.4.4 Resize text errors (N = 4,436) comprised 10.1% of all errors in this study. These errors pertain to the size and position of text on a webpage. Nearly all resize text errors in this study arose from font that was too small on the webpage for web users with low vision. In some instances, the font on webpages was at size-8 or smaller, making it very difficult for web users with low vision to read the text. If web users must zoom in to read the font, other content on the webpage may become distorted, leading to confusion. Often, markup language will default to a certain size font unless the web developer specifies a preferred font size for the entire webpage.

## *2. Suggested solutions*

Many of the Level-A 1.3.1 errors in this study could be fixed by adopting three approaches. First, unbolding text and writing text in lowercase would improve the readability of text on the webpage (as shown in Figure 1). Second, shortening the text description of non-text web elements would increase the intelligibility of the content and non-text web elements on the webpage. Third, providing a more informative description of non-text web elements would allow users to learn more about the webpage and be better able to navigate the webpage's content. Many of this study's Level-A 1.1.1 errors could be fixed by entering unique and descriptive alternative text and metadata into each web element to ensure that students with disabilities can discern between web elements and access all non-text content on a webpage. Level-A 1.4.4 errors could be fixed by increasing the size of webpage font.

## **B. Frequent operability errors and suggested corrections**

*1. Common errors*

Level-A 2.4.4 Link purpose web errors (N = 6,514) comprised 14.7% of all web accessibility errors in this study. Many of these errors were identified because a hyperlink lacked a description in its metadata, meaning that web users could not hover a mouse over the hyperlink and have its description read aloud. Students with disabilities using screen reader technology, such as Voiceover, use this metadata to understand a hyperlink's purpose. Level-A 2.4.4 errors were also identified for redundant hyperlink destinations with different hyperlink descriptive text, meaning hyperlinks with different descriptions lead to the same webpage. Redundant hyperlink destinations may be confusing for students with disabilities, as a web user may expect that hyperlinks with different descriptive text lead to different webpages. Finally, many other Level-A 2.4.4 errors identified uninformative hyperlink descriptions, including descriptions such as "link" and "link to page." These descriptions are not informative enough for students with disabilities. For example, a student with low-vision may hover their mouse over a hyperlink that visually reads "Undergraduate Application Instructions" because they want to apply to the institution, but the hyperlink may be read by screen readers as "link to page," which would not tell the student that they located the correct hyperlink to apply to the institution. Instead, content editors should provide rich descriptions of hyperlinks (e.g., "link to physics curriculum webpage" instead of "link"), so that students with disabilities are given enough information to successfully navigate the webpage and find the content they need.

*2. Suggested solutions*

Many of the Level-A 2.4.4 errors could be resolved by making sure that all hyperlinks include unique, informative descriptions and that multiple hyperlinks with differing descriptions do not lead to the same webpage.

**C. Frequent robustness errors and suggested corrections**

*1. Common error*

Robust Level-A 4.1.2 Name, role, value errors were abundant (N = 3,498) and comprised 7.9% of all web accessibility errors in this study. Level-A 4.1.2 errors pertain to whether the hyperlinks on a webpage contain informative href attributes. Href stands for "hypertext reference" and is information that specifies the URL of the webpage that the hyperlink goes to. In addition, many Level-A 4.1.2 errors identified incomplete Accessible Rich Internet Application (ARIA) attributes. ARIA attributes are additional pieces of information that communicate with assistive technologies, informing the user about what kind of web elements are on the webpage and how to interact with the elements. Both href and ARIA attributes of hyperlinks are important when a wide range of assistive technologies attempt to communicate with web elements on a webpage. For example, if a hyperlink is missing an href attribute, the hyperlink will not direct the web user to a different webpage; the hyperlink will be dead. Additionally, if ARIA attributes are present in web elements on the middle of the webpage but missing on web elements at the top of the page, an assistive technology may not be able to recognize the web element and describe the element to a user.

*2. Suggested solution*

Level-A 4.1.2 errors require more extensive knowledge of markup language (e.g., HTML, Java), so they should be addressed by web administrators and developers working at institutions of higher education to ensure that each institutional webpage includes the most robust and informative information to allow the widest range of assistive technologies access to the content. Web administrators should be aware of what original content has been published on the webpage, and which web elements are present in other areas of the webpage. As Level-A 4.1.2 errors primarily address original content developed by the web administrator, these errors are most prevalent when proprietary web elements are developed but under described, such as the case of missing href or ARIA attributes. It takes more technical skill to accessibly generate complicated web elements; thus, physics departments should strive to create webpages using simple web elements.

**D. Other errors with simple solutions**

One perceivable Level-A 1.2.2 captions (prerecorded) error was identified in this study. This error indicates that a video was not captioned, making it difficult for deaf web users to access the content. All video content should be captioned.

Understandable Level-A 3.1.1 language of page errors (<1% of all errors) indicate that the spoken language of the webpage (i.e., "en" for English) was not included in the language attributes of the web element. In this case, students with disabilities who are English language learners may not be able to access information on webpages that are either not translated or do not specify the language of the webpage. The language used on a webpage should be identified in the attributes of the web element.

Operable Level-A 2.1.1 keyboard errors (1.4% of all errors) indicate that many web elements were not written in ways that allow keyboard-centric assistive technologies access to the content. Although Level-A 2.1.1 errors involve many different aspects of web accessibility, it is important to note that keyboard-centric assistive technologies are widely used by people with disabilities when accessing webpage. Physics curriculum writers and institutional web developers should pay close attention to whether their webpages are robust enough to allow keyboard-centric assistive technologies access to the content. Professionals interested in improving the web accessibility of their webpages could look to WebAIM's website, published specifically to educate people who wish to improve their web accessibility and provide more inclusive and robust online information for people with disabilities.

**E. Limitations**

This case study focused only on U.S. webpages related to undergraduate physics curriculum and graduate research opportunities and our sample did not include any private, for-profit institutions. Moreover, this study was limited by the evaluation of web accessibility using a single accessibility audit software and one assistive technology. Given the time intensive nature of data collection and analysis, the research team decided it was only feasible to evaluate the webpages using one audit software and one assistive technology, understanding that webpages often change on a daily or hourly basis. In addition, this study only analyzed webpages and not other forms of media, such as PDF files or PowerPoint presentations.

As a result, future research could expand upon our sample size, use a greater number of accessibility audit software programs and assistive technologies, and employ a larger research team to provide a more comprehensive picture of the web accessibility of physics curriculum webpages at U.S. institutions of higher education. Additionally, future research could address other webpages students must interact with to navigate postsecondary education, such as financial aid, student affairs, and Title-IX webpages.

## V. IMPLICATIONS

Below we explicate the implications of our study for four groups of stakeholders: physics students, instructors, university administrators, and the broader physics community.

### A. Current and prospective physics students with disabilities

Current and prospective physics students with disabilities need to be aware that physics webpages are inaccessible and can create barriers to their participation in physics. As such, students should be ready to advocate for themselves to get access to information they need. They should also be open to communicating directly with institutional faculty and staff. The Washington Disabilities, Opportunities, Internetworking, and Technology (DO-IT) network provides a myriad of resources for students with disabilities.

### B. Physics instructors and staff

Depending on institutional norms, physics instructors and support staff may be responsible for creating web content and editing course webpages via a learning management system and/or may collaborate with information technology staff on institutional and departmental webpages. In either case, physics instructors and staff should be aware of the technological hurdles that students with disabilities face when accessing online content and should explore alternative ways of delivering content to students with disabilities to ensure that all students have equal access to physics curricula and other learning materials. Instructors and staff should also be aware that students with disabilities will need to contact them individually to get access to relevant information and should be ready and eager to provide this information to any student who asks for it. Bradbard and Peters (2010) provide an introduction to web accessibility for faculty, and Amundson (2009) provides five steps for instructors to increase the accessibility of their webpages.

McGough (2016) found that institutions will not make changes to make their webpages more accessible unless there are outside pressures, such as lawsuits. Thus, faculty should also push university administration to proactively create webpages and content that are accessible to all students.

**C. University administrators and institutional leaders**

Advances in technology have rendered the Internet and postsecondary webpages essential resources for all educational stakeholders, including students. However, advances in technology bring challenges when crafting online content that is truly accessible to all students, not just those without physical, developmental, or cognitive limitations.

Institutions of higher education have faced hundreds of disability-related lawsuits brought by people who were not able to have access to equal educational opportunity online. If institutional leaders want to support all students and increase access to their institution, web accessibility must be prioritized. Moreover, research tells us that students with disabilities are underrepresented in STEM degree programs and the STEM workforce (NSF, 2019). Improvements in web accessibility would not only help avoid costly litigation, but more importantly would increase access to STEM major information for students with disabilities. Also, if an institution's webpages are noticeably more accessible than another institution's, then students with disabilities would be more likely to enroll at the more accessible institution (Burdett, 2013; Daun-Barnett & Das, 2013). Burgstahler (2006) and Tandy and Meacham (2009) provide suggestions for administrators on how to increase the digital accessibility of their institution's webpages. Individuals can investigate the accessibility of webpages using WebAIM's WAVE tool ([https://webaim.org/](https://webaim.org/)).

**D. Broader physics community**

The broader physics community needs to recognize that we have a significant problem with the digital accessibility of our webpages. If we want to increase the representation of people with disabilities in physics and, more broadly, the diversity of the physics community, then we need to increase accessibility. Creating accessible web content should be a priority for the physics community. If we do not create accessible webpages, we send the message that we do not expect people with disabilities to participate in our community. We must continue to include disability as a dimension of diversity that we care about.

To increase the accessibility of physics webpages, we need to provide instructors, faculty, staff, and non-profit leaders with support (financial, intellectual, and moral) to press universities and organizations to make changes toward accessibility. For example, professional societies could maintain "tips and tricks" for creating accessible web content and could include accessibility checks in recommendations for program review.

**Abbreviations:** ARIA: Accessibility Rich Internet Application; HREF: Hypertext Reference; IPEDS: Integrated Postsecondary Education Data System; NSF: National Science Foundation; STEM (Science, Technology, Engineering, and Mathematics); U.S.: United States of American; W3C: World Wide Web Consortium; WCAG: Web Content Accessibility Guidelines

**Declarations:**

Availability of data and materials: The dataset supporting the conclusions of this article are available from the authors upon request.

Competing interests: The authors declare that they have no competing interests.

Funding: This work is supported in part by National Science Foundation award # 1750515.

Authors' contributions: Z.W.T., E.S., and J.C.C. coordinated to conceive of and conduct this study. Z.W.T. conducted the Tenon and Voiceover analysis while E.S. interpreted the findings for the STEM community. All authors contributed to writing, read and approved the final manuscript.

Appendix

Table A1: Web Content Accessibility Guidelines (WCAG) principles, guidelines, and examples of success criteria and their levels (Note: this is not an exhaustive list).

| **Principles** | **Guidelines** | **Example Success Criteria** |
|---|---|---|
| 1. **Perceivable** information and user interface | 1.1 Text alternatives for non-text content | *1.1.1 Non-text Content*: All non-text content that is presented to the user has a text alternative that serves the equivalent purpose except for a short list of exceptions with individual success criteria (e.g., If non-text content is a control or accepts user input, then it has a name that describes its purpose.) (**Level A**) |
| | 1.2 Captions and other alternatives for multimedia | *1.2.2 Captions (Prerecorded):* Captions are provided for all pre-recorded audio content in synchronized media, except when the media is a media alternative for text and is clearly labeled as such. **(Level A)** <br> *1.2.5 Audio Description (Prerecorded):* Audio description is provided for all prerecorded video content in synchronized media. **(Level AA)** <br> *1.2.6 Sign Language (Prerecorded):* Sign language interpretation is provided for all prerecorded audio content in synchronized media. **(Level AAA)** |
| | 1.3 Adaptable: Content can be presented in different ways without losing information or structure | *1.3.1 Info and Relationships:* Information, structure, and relationships conveyed through presentation can be programmatically determined or are available as text **(Level A)** <br> *1.3.2 Meaningful Sequence:* When the sequence in which content is presented affects its meaning, a correct reading sequence can be programmatically determined. **(Level A)** <br> *1.3.3 Sensory Characteristics:* Instructions provided for understanding and operating content do not rely solely on sensory characteristics of components such as shape, size, visual location, orientation, or sound. **(Level A)** |
| | 1.4 Distinguishable: Content is easier to see and hear | *1.4.1 Use of Color:* Color is not used as the only visual means of conveying information, indicating an action, prompting a response, or distinguishing a visual element. **(Level A)** <br> *1.4.3 Contrast (Minimum):* The visual presentation of text and images of text has a contrast of at least 4.5:1, except for large-scale text, incidental text (e.g., pure decoration), and logotypes. **(Level AA)** <br> *1.4.4 Resize Text:* Except for captions and images of text, text can be resized without assistive technology up to 200 percent without loss of content or functionality. **(Level AA)** <br> *1.4.9 Images of Text (No Exception):* Images of text are only used for pure decoration or where a particular presentation of text is essential to the information being conveyed. **(Level AAA)** |

| 2. **Operable** user interface and navigation | 2.1 Keyboard Accessible: Functionality is available from a keyboard | *2.1.1 Keyboard:* All functionality of the content that is operable through a keyboard interface without requiring specific timings for individual keystrokes, except where the underlying function requires input that depends on the path of the user's movement and not just the end points. (**Level A**) <br> *2.1.2 No Keyboard Trap:* If keyboard focus can be moved to a component of the page using a keyboard interface, then focus can be moved away from that component using only a keyboard interface, and, if it requires more than unmodified arrow or tab keys or other standard exit methods, the user is advised of the method for moving focus away. (**Level A**) |
|---|---|---|
| | 2.2 Enough Time: Users have enough time to read and use the content | *2.2.2 Pause, Stop, Hide:* For moving, blinking, scrolling, or auto-updating information, all of the following are true) 1) starts automatically, (2) lasts more than five seconds, and (3) is presented in parallel with other content, there is a mechanism for the user to pause, stop, or hide it unless the movement, blinking, or scrolling is part of an activity where it is essential. (**Level A**) <br> *2.2.5 Re-authenticating:* When an authenticated session expires, the user can continue the activity without loss of data after re-authenticating. (**Level AAA**) |
| | 2.3 Seizures: Content does not cause seizures | *2.3.1 Three Flashes or Below Threshold:* Webpages do not contain anything that flashes more than three times in any one second period, or the flash is below the general flash and red flash thresholds. (**Level A**) <br> *2.3.2 Three Flashes:* Webpages do not contain anything that flashes more than three times in any one second period. (**Level AAA**) |

| | 2.4 Navigable: Users can easily navigate, find content, and determine where they are on a webpage | *2.4.1 Bypass Blocks:* A mechanism is available to bypass blocks of content that are repeated on multiple webpages. (**Level A**) <br> *2.4.2 Page Titled:* Webpages have titles that describe topic or purpose. (**Level A**) <br> *2.4.3 Focus Order:* If a webpage can be navigated sequentially and the navigation sequences affect meaning or operation, focusable components receive focus in an order that preserves meaning and operability. (**Level A**) <br> *2.4.4 Link Purpose (In Context):* The purpose of each link can be determined from the link text alone or from the link text together with its programmatically determined link context, except where the purpose of the link would be ambiguous to users in general. (**Level A**) <br> *2.4.6 Headings and Labels*: Headings and labels describe topic or purpose. (**Level AA**) |
|---|---|---|
| 3. **Understandable** information and user interface | 3.1 Readable: Text is readable and understandable | *3.1.1 Language of Page:* The default human language of each webpage can be programmatically determined. (**Level A**) <br> *3.1.3 Unusual Words:* A mechanism is available for identifying specific definitions of words or phrases used in an unusual or restricted way, including idioms and jargon. (**Level AAA**) <br> *3.1.4 Abbreviations:* A mechanism for identifying the expanded form of meaning of abbreviations is available. (**Level AAA**) |
| | 3.2 Predictable: Content appears and operates in predictable ways | *3.2.1 On Focus:* When any component receives focus, it does not initiate a change of context. (**Level A**) <br> *3.2.2 On Input:* Changing the setting of any user interface component does not automatically cause a change of context unless the user has been advised of the behavior before using the component. (**Level A**) <br> *3.2.4 Consistent Identification:* Components that have the same functionality within a set of webpages are identified consistently. (**Level AA**) |

| | 3.3 Input Assistance: Users are helped to avoid and correct mistakes | *3.3.1 Error Identification:* If an input error is automatically detected, the item that is in error is identified and the error is described to the user in text. (**Level A**)<br>*3.3.2 Labels or Instructions:* Labels or instructions are provided when content requires user input. (**Level A**)<br>*3.3.3 Error Suggestion:* If an input error is automatically detected and suggestions for the correction are known, then the suggestions are provided to the user, unless it would jeopardize the security or purpose of the content. (**Level AA**)<br>*3.3.6 Error Prevention (All):* For Webpages that require the user to submit information, at least of the following is true: 1. Reversible: Submissions are reversible.; 2. Checked: Data entered by the user is checked for input errors and the uses is provided an opportunity to correct them.; 3) Confirmed: A mechanism is available for reviewing, confirming, and correcting information before finalizing the submission. (**Level AAA**) |
|---|---|---|
| 4. **Robust** content and reliable information | 4.1 Compatible: Content is compatible with current and future user tools | *4.1.1. Parsing:* In content implemented using markup languages, elements have complete start and end tags, elements are nested according to their specifications, elements do not contain duplicate attributes, and any IDs are unique, except where the specifications allow these features. (**Level A**)<br>*4.1.2 Name, Role, Value:* For all user interface components, the name and role can be programmatically determined; states, properties, and values that can be set by the user can be programmatically set; and notification of changes to these items is available to user agents, including assistive technologies. (**Level A**) |